\documentclass[runningheads,a4paper]{llncs}

\usepackage{amssymb}
\usepackage{graphicx}
\usepackage{url}
\urldef{\mailsa}\path|amaatouq@mit.edu|
\newcommand{\keywords}[1]{\par\addvspace\baselineskip
\noindent\keywordname\enspace\ignorespaces#1}

\usepackage{color}
\definecolor{Blue}{rgb}{0,0,1}

\definecolor{Orange}{rgb}{1,0.5,0}

\definecolor{Green}{rgb}{0,1,0}

\begin{document}

\mainmatter  

\title{The Role of Reciprocity and Directionality of Friendship Ties in Promoting Behavioral Change}


\titlerunning{Directionality of Friendship Ties and Behavioral Change}

%
%
\author{Abdullah Almaatouq\thanks{corresponding authors}%
\and Laura Radaelli\and Alex Pentland \and Erez Shmueli}
%

\authorrunning{Almaatouq et. al.}

\institute{Massachusetts Institute of Technology,\\
77 Massachusetts Ave, Cambridge, MA 02139, USA\\
\mailsa\\
}

%
%

\toctitle{Lecture Notes in Computer Science}
\tocauthor{Authors' Instructions}
\maketitle

\begin{abstract}
Friendship is a fundamental characteristic of human beings and usually assumed to be reciprocal in nature. Despite this common expectation, in reality, not all friendships by default are reciprocal nor created equal. Here, we show that reciprocated friendships are more intimate and they are substantially different from those that are not. We examine the role of reciprocal ties in inducing more effective peer pressure in a cooperative arrangements setting and find that the directionality of friendship ties can significantly limit the ability to persuade others to act. Specifically, we observe a higher behavioral change and more effective peer-influence when subjects shared reciprocal ties with their peers compared to sharing unilateral ones. Moreover, through spreading process simulation, we find that although unilateral ties diffuse behaviors across communities, reciprocal ties play more important role at the early stages of the diffusion process.
\end{abstract}

\section{Introduction}

Friendship is a fundamental characteristic of human relationships and individuals generally presume it to be reciprocal in nature. Despite this common expectation~\cite{Willard93}, in reality not all friendships are reciprocal~\cite{laursen1993adolescents,hartup1993adolescents}. The implications of friendships on an individual's behavior depend as much on the identity of his friends as on the quality of friendships~\cite{hartup1996company}. Among qualities of a relationship, reciprocity can substantially differentiate a friendship from many others.
It is reasonable to think that relationships that are reciprocated are substantially different from those that are not~\cite{hartup1996company}.

Moreover, in recent years, peer-support programs are emerging as highly effective and empowering ways to leverage peer influence to support behavioral change~\cite{ford2013systematic}.
One specific type of peer-support programs is the ``buddy system,'' in which individuals are paired with another person (i.e., a buddy) who has the responsibility to support their attempt to change their behavior.
Such a system has been used to shape people's behavior in various domains including smoking cessation~\cite{may2000social}, weight loss~\cite{stock2007healthy}, diabetes management or alcohol misuse~\cite{rotheram2012diabetes}. 

Consequently, the need to understand the factors that impact the level of influence individuals exert on one another is of great practical importance.
Recent studies have investigated how the effectiveness of peer influence is affected by different social and structural network properties, such as clustering of ties~\cite{centola2010spread}, similarity between social contacts~\cite{centola2011experimental}, and the strength of ties~\cite{aral2013tie}. However, how the effectiveness of social influence is affected by the reciprocity and directionality of friendship ties is still poorly understood.

When analyzing self-reported relationship surveys from several experiments, we find that only about half of the friendships are reciprocal.
These findings suggest a profound inability of people to perceive friendship reciprocity, perhaps because the possibility of non-reciprocal friendship challenges one's self-image. 

We further show that the asymmetry in friendship relationships has a large effect on the ability of an individual to persuade others to change their behavior. 
Moreover, we show that the effect of directionality is larger than the effect of the self-reported strength of a friendship tie~\cite{aral2013tie} and thus of the implied `social capital' of a relationship. 
Our experimental evidence comes through analysis of a fitness and physical activity intervention, in which subjects were exposed to different peer pressure mechanisms, and physical activity information was collected passively by smartphones.
In this experiment, we find that effective behavioral change occurs when subjects share reciprocal ties, or when a unilateral friendship tie exists from the person applying the peer pressure to the subject receiving the pressure, but not when the friendship tie is from the subject to the person applying peer pressure. 

Our findings suggest that misperception of friendships' character for the majority of people may result in misallocation of efforts when trying to promote a behavioral change.


\section{Results}

\subsection{Reciprocity and Intimacy}

Despite the unique characteristics and importance of reciprocal friendships, reciprocity is implicitly assumed in very many scientific studies of friendship networks: in their analysis they either mark two individuals as friends of each other, or as not being friends.
However, not all friendships are reciprocal, as we proceed to demonstrate.

We analyze surveys that were used to determine the closeness of relationships (i.e., friendships) among participants in the Friends and Family study.
Each participant in the study scored other participants on a $0-7$ scale, where a score of $0$ meant that the participant was not familiar with the other, and $7$ that the participant was very close to the other.

\begin{figure}[ht]
	\centering
	\includegraphics[width=1\columnwidth]{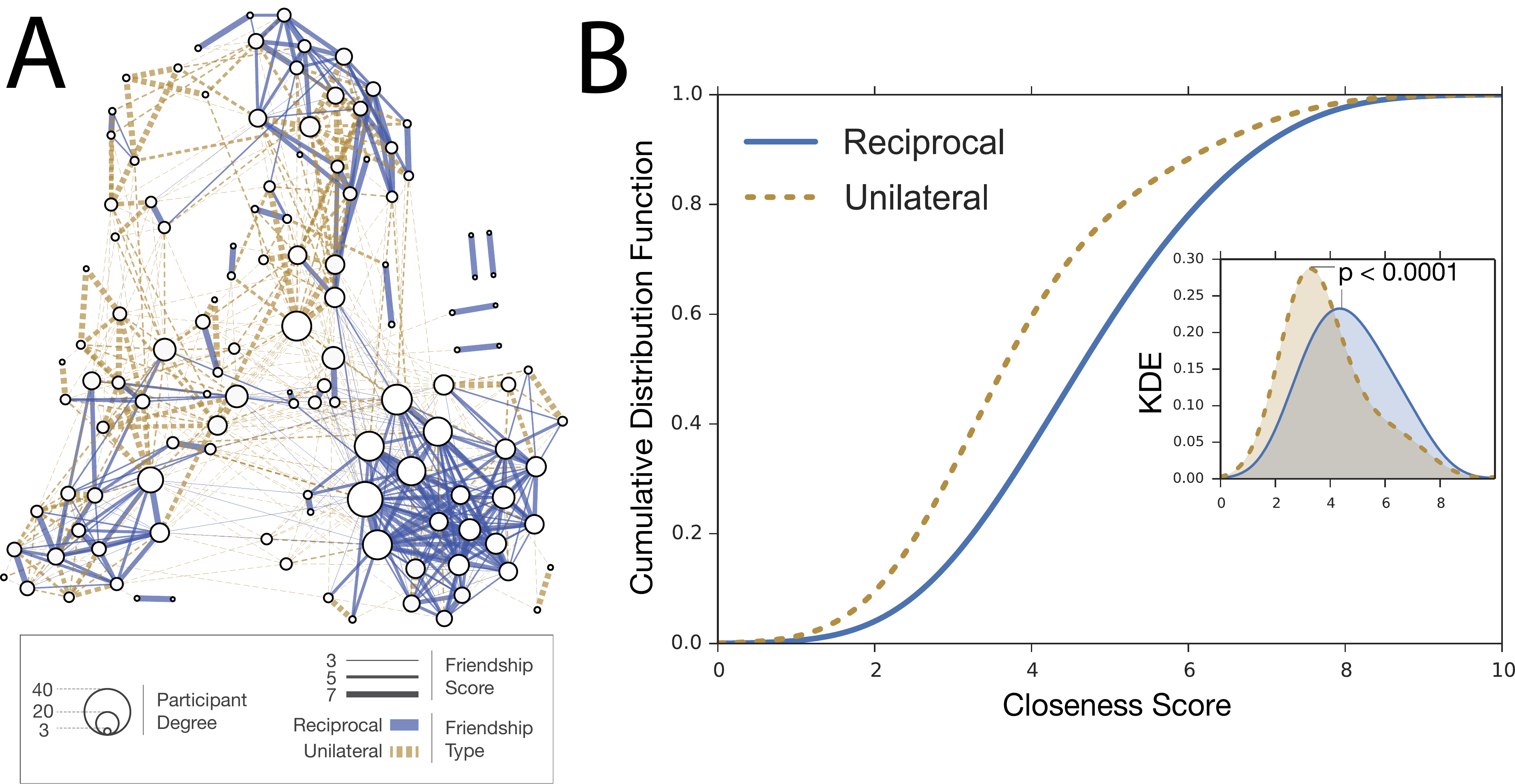}
	\caption{\textbf{Subfigure~A} depicts the undirected friendship nomination graph in the Friends and Family study, where nodes represent participants and edges represent friendship ties. \textbf{Subfigure~B} shows the distribution of closeness scores for reciprocal and unilateral ties. ECDF and KDE of closeness scores are computed separately for unilateral ties (dashed line) and reciprocal ties (solid line).}
	\label{fig:closeness_scores}
\end{figure}

The self-reported closeness scores were then used to build the friendship network. 
Similar to~\cite{aharony2011social}, we considered only explicit friendship ties ($closeness > 2$). In this network, we consider a friendship tie to be ``reciprocal'' when both participants identify each other as friends. Alternatively, the tie is ``unilateral'' when only one of the participants identifies the other as a friend. 
Figure~\ref{fig:closeness_scores} depicts the resulting network which consists of 122 nodes and 698 edges (i.e., explicit friendships), of which 315 are reciprocal (i.e., 45\%) and 383 are unilateral (i.e., 55\%).
Surprisingly, more than half of the participants' friendship ties are not reciprocated, which indicates the non-intuitive observation that people are very vulnerable to misjudging their friendship relationships and implies that people are unable to perceive reciprocity~\cite{almaatouq2006}.

We find this result to be consistent across many self-reported friendship networks that we have analyzed: only 45\% (315 out of 698) of friendships are reciprocal in the Friends and Family dataset~\cite{aharony2011social}, 34\% (28 out of 82) in the Reality Mining dataset~\cite{eagle2006reality}, 35\% (555 out of 1596) in the Social Evolution dataset~\cite{madan2012sensing}, and 49\% (102 out of 208) in the Strongest Ties dataset~\cite{de2014strength}.
The first three surveys were collected at an American university, and the fourth at a European university.

Similarly, a previous study~\cite{Vaquera2008} in which adolescents were asked to nominate at most 10 of their best school friends (5 male and 5 female) found that only 64\% of the reported friendships were indeed reciprocal.
Our findings reinforce this finding by investigating multiple datasets from two continents, and by using complete nomination networks (in which each participant is asked about every other participant), resulting in an even more prominent lack of reciprocity.

Finally, analyzing the closeness scores associated with the two types of ties in the Friends and Family friendship network reveals that participants that share a reciprocal friendship tend to score each other higher (on average) when compared to participants that share unilateral friendship.
More specifically, the average closeness score of reciprocal ties ($4.7$) is almost one point higher than the average score of unilateral ties ($3.9$) and the difference is statistically significant (two-sample T-test $p<0.0001$).

\subsection{Induced peer pressure}
Social scientists have long suspected that reciprocal friendships are more intimate, provide higher emotional support, and form a superior resource of social capital when compared to those that are not reciprocated. This holds whether or not any party of the dyad is aware of the status of reciprocity embedded in their relationships~\cite{Vaquera2008}. However, we hypothesize that `reciprocity' and `directionality' of friendships may be critical factors in promoting peer influence, beyond the mere effect of the total tie `strength' bound up in the relationship. 

To support our hypothesis, we investigate the FunFit experiment -- a fitness and physical activity experimental intervention -- conducted within the Friends and Family study population during October to December of 2010.
The experiment was presented to participants as a wellness game to help them increase their daily activity levels.  Subjects received an `activity app' for their mobile phone which passively collected their physical activity data and showed the participants how their activity level had changed relative to their previous activity level, and the amount of money they had earned by being more active.
108 out of the 123 active Friends and Family subjects at that time elected to participate and were allocated into three experimental conditions, allowing us to isolate different incentive mechanisms varying monetary reward, the value of social information, and social pressure/influence:
\begin{itemize}
	\item \textbf{Control:} subjects were shown their own progress and were given a monetary reward based on their own progress in increasing physical activity relative to the previous week.
	\item \textbf{Peer See:} subjects were shown their own progress and the progress of two ``buddies‚'' in the same experimental group, and were given a monetary reward based on their own progress in increasing physical activity relative to the previous week.
	\item \textbf{Peer Reward:} subjects were shown their own progress and the progress of two ``buddies‚'' in the same experimental group, but their rewards depended only on the progress of the two ``buddies''. This condition realizes a social mechanism based on inducing peer-to-peer interactions and peer pressure~\cite{mani2013inducing}.
\end{itemize}
However, for the purpose of our analysis in this section, we combine the samples from the two peer pressure treatments, as we are interested in peer pressure regardless of the incentive structure, and omit the control group.

During the initial 23 days of the experiment (Oct 5 - Oct 27), denoted as P1, the baseline activity levels of the subjects were collected. The actual intervention period is denoted as P2.
During the intervention period, the subjects were given feedback on their performance in the form of a monetary reward.
The monetary reward was calculated as a function of the subject's activity data relative to the previous week and was divided according to the subject's experimental condition (i.e., Peer See and Peer Reward).
Note that the physical activity was measured passively by logging the smartphone accelerometer (as opposed to self-reported surveys) and the game was not designed as a competition, every subject had the potential to earn the maximal reward. A previously non-active participant could gain the same reward as a highly active one, while the highly active person would need to work harder.

The results in~\cite{aharony2011social} show that the two social conditions (i.e. Peer See and Peer Reward) do significantly better than the control group.
Furthermore, the results suggest that there is a complex contagion effect~\cite{centola2007complex}, due to the reinforcement of the behavior from multiple social contacts~\cite{centola2010spread,centola2007complex}, related to pre-existing social ties between participants.
Our analysis here focuses on the role of reciprocity and directionality of friendship ties in this contagion process.

In order to investigate the role of reciprocity and directionality of friendship ties in the contagion process, we performed a regression analysis in which the dependent variable was the change in physical activity between the post-intervention phase and the pre-intervention phase (i.e., the average daily physical activity in P2 divided by the average daily physical activity in P1).

For our study, we refer to a participant whose behavior is being analyzed as ``ego'', and participants connected to the ego (i.e., experimental ``buddies'') are referred to as ``alters''.
Because friendship nominations are directional, we studied the three possible types of friendships (from the prospective of the ego) as independent variables:
an ``ego perceived friend'', in which an alter identifies an ego as a friend (i.e., incoming tie);
an ``alter perceived friend'' in which an ego identifies an alter as a friend (i.e., outgoing tie);
and a ``reciprocal friend,'' in which the identification is bidirectional (i.e., reciprocal tie).
Finally, we also included the tie strength (i.e., the sum of the closeness scores between an ego and his or her alters) as a control variable, which has been previously investigated as a moderator of the effect of social influence~\cite{aral2013tie}.

\begin{figure}[ht]
	\centering
	\includegraphics[width=0.75\columnwidth]{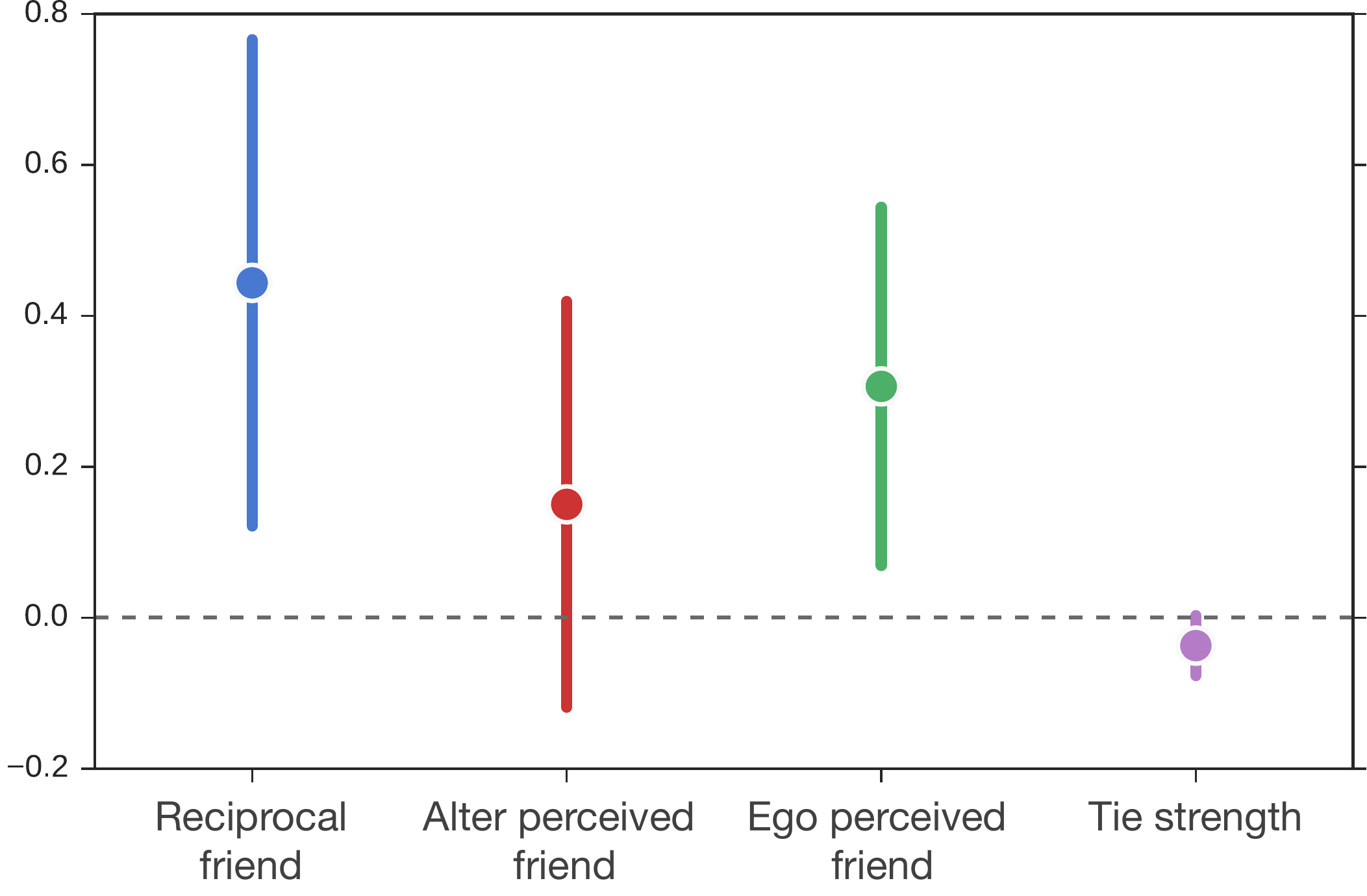}
	\caption{Change in physical activity under experiment conditions shows that the type of friendship is relevant to the effectiveness of the induced peer pressure. The plot shows the mean effect size of the covariates (solid circles) and the 95\% confidence intervals (bars).}
	\label{fig:peer_pressure}
\end{figure}

Figure~\ref{fig:peer_pressure} reports the effects found in our regression analysis (recall that the dependent variable in our model is the change in activity for the egos).
We find that the reciprocity and directionality of a friendship have an effect on the amount of induced peer pressure, and these effects are much larger than the total tie strength.

	The strongest effect for both treatment groups ($N = 76$) in this study was found for the reciprocal factor ($p < 0.01$) even when controlling over the strength of the tie (the tie strength is weakly significant $p = 0.07$).
	That is, alters in reciprocal friendships have more of an effect on the ego than alters in other types of friendships.
	
	Interestingly, when the ego was perceived as a friend by the alters (i.e., incoming edges from the alters to the ego), the effect was also found to be positive and significant ($p < 0.05$).
	On the other hand, no statistically significant effect was found when the alters were perceived as friends by the ego (i.e., outgoing edges from the ego to the alters).
	Therefore, the amount of influence exerted by individuals on their peers in unilateral friendship ties seems to be dependent on the direction of the friendship.
	
	Unlike previous works on social contagion effects~\cite{christakis2007spread,fowler2009dynamic}, which were conducted without peer-to-peer incentives, we find that influence does not flow from nominated alter to nominating ego. Surprisingly, alter's perception of ego as a friend would increase alter's ability to influence ego's behavior when ego does not reciprocate the friendship.
	We attribute this difference to the fact that there is a peer-to-peer incentive mechanism, and therefore there are likely to be differences in communication when the alters believe the ego to be their friend versus when they do not.

\subsection{Reciprocity and Global Adoption}
\label{sec:global_adoption}
In order to understand the effect of reciprocal ties on global behavior adoption, we experimented with a variation of the classic epidemic spreading model, Susceptive-Infected (SI) model.
We refer to this variant as the Bi-Directional Susceptive-Infected (BDSI) model.
Unlike the classic SI in which behavior is transmitted along edges with a constant probability, the proposed BDSI model considers the direction in which behaviors can be transmitted with different probabilities based on the direction and type of edges -- i.e. $p_{rec}$ for reciprocal edges and $p_{+}$/$p_{-}$ for the two possible directions of unilateral edges.

\begin{figure}[h!]
	\centering
	\includegraphics[width=0.75\columnwidth]{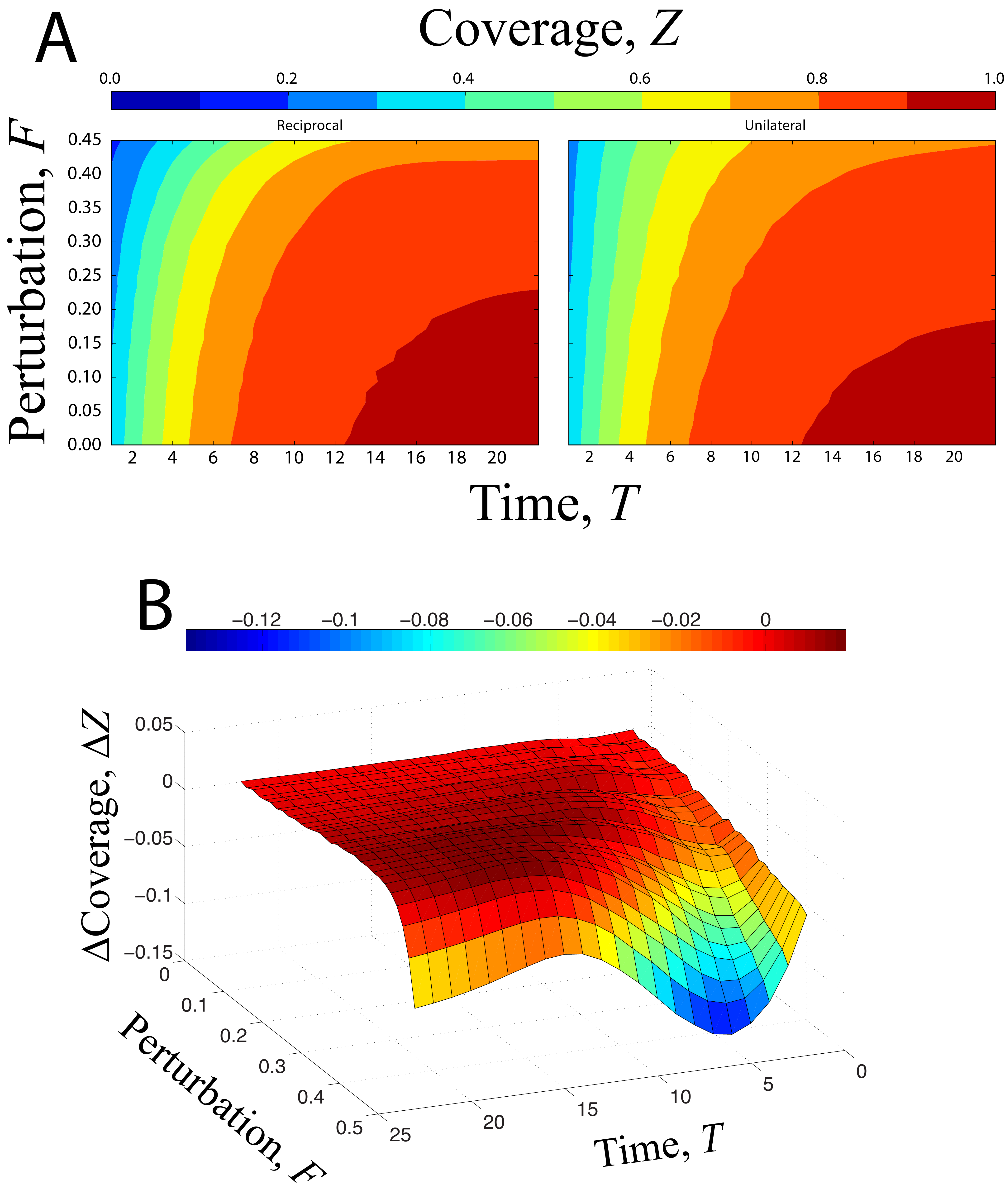}
	\caption{
		\textbf{Subfigure~A} demonstrates the effect of perturbation on the coverage and speed of adoption. \textbf{Subfigure~B} illustrates how the coverage and speed decay faster when removing reciprocal edges in comparison with unilateral edges.}
	\label{fig:simulation}
\end{figure}

In order to observe the effects of reciprocal edges on diffusion, we employ an edge percolation process in which we measure the coverage (i.e., number of infected nodes), denoted by $Z$, and time to infect, denoted by $T$, when removing reciprocal and unilateral edges successively (i.e., perturbation $F$).
That is, $F$ is the percentage of edges removed in perturbation.
We find the nature of the simulation results are qualitatively independent of the choice of $p_{rec}$, $p_{+}$ and $p_{-}$ given that $p_{rec} \ge p_{+} \ge p_{-}$.

Figure~\ref{fig:simulation} shows the behavior adoption coverage when simulating the BDSI model on the self-reported friendship network from the Friends and Family dataset.
As can be seen in the figure, the coverage $Z$ decays much faster when removing reciprocal edges (left figure) compared with removing the same amount of unilateral edges (right figure).
Moreover, the difference in coverage $\Delta Z$ is affected remarkably by the removal of reciprocal edges most notably in the early stages of the diffusion process (e.g., $T\in [5,10]$). 
This can be attributed to the rapid diffusion within a single community through reciprocal edges, corresponding to fast increases in the number of infected users in early stages of the diffusion process, followed by plateaus, corresponding to time intervals during which no new nodes are infected the behavior escapes the community (i.e., through the strength of weak tie~\cite{granovetter1973strength}) to the rest of the network through unilateral edges.

\section{Future Work \& Discussion}
\label{sec:conclusion}
In this paper we have demonstrated the important role that  reciprocity and directionality of friendship ties play in inducing effective social persuasion.
We have also shown that the majority of individuals have difficulty in judging the reciprocity and directionality of their friendship ties (i.e., how others perceive them), and that this can be a major limiting factor for the success of cooperative arrangements such as peer-support programs. Finally, through spreading process simulation, the experimental results highlight the important role that reciprocal ties play in the spreading of behaviors at the early stages of the process.

Previous studies have found that people tend to adopt the behaviors of peers that they are passively exposed to, with the explicit self-reported friends and intimate social acquaintances playing a peripheral role (e.g.,~\cite{madan2010social,centola2010spread}).
Other studies have shown that passive exposures to peer behavior can increase the chances of becoming obese~\cite{christakis2007spread,madan2010social}, registering for a health forum Web site~\cite{centola2010spread}, signing up for an Internet-based diet diary~\cite{centola2011experimental}, or adopting computer applications~\cite{aral2012identifying}.
However, our results suggest a fundamental difference between how social learning (i.e., passive exposure) and social persuasion (i.e., active engagement) spread behaviors from one person to another.

The findings of this paper have significant consequences for designing interventions that seek to harness social influence for collective action.  This paper also has significant implications for research into peer pressure, social influence, and information diffusion as these studies have typically assumed undirected (reciprocal) friendship networks, and may have missed the role that the directionality of friendship ties plays in social influence.

\bibliographystyle{splncs}

\end{document}